\documentclass[aps,preprint,nofootinbib]{revtex4}
\usepackage{amsfonts}
\usepackage{amssymb}

\usepackage{rotating}
\newcommand{\be}{\begin{equation}}
\newcommand{\ee}{\end{equation}}
\newcommand{\bea}{\begin{eqnarray}}
\newcommand{\eea}{\end{eqnarray}}
\begin{document}
\title{On extracting information about hadron-nuclear interaction from
hadronic atom level shifts}

\author{J. R\'{e}vai\footnote{Corresponding author:
revai@rmki.kfki.hu}} \affiliation{Research Institute for Particle
and Nuclear Physics, H-1525 Budapest, P.O.B. 49, Hungary}

\author{N.V. Shevchenko}
\affiliation{Nuclear Physics Institute, 25068 \v{R}e\v{z}, Czech
Republic}

\date{\today}





\begin{abstract}
It is argued, that adjusting strong potentials directly to
observed hadronic atom level shifts may lead to significantly
different scattering lengths, than those, predicted by the Deser
formula~\cite{Des}.
On the example of the 1s level shift of kaonic hydrogen it is
demonstrated, that the usually adopted Deser values deduced from
the two recent measurements in KEK~\cite{KEK} and by the DEAR
Collaboration~\cite{DEAR} $a_D({\rm KEK}) = 0.78 - 0.49i$ fm and
$a_D({\rm DEAR}) = 0.47 - 0.3i$ fm should be replaced by
$a_s({\rm KEK})\simeq 0.85 - 0.62i$ fm and
$a_s({\rm DEAR})\simeq 0.49 - 0.35i$ fm, correspondingly.
\end{abstract}

\maketitle

\section{Introduction}
One of the most relevant sources of information on hadron-nucleus
interaction is the measurement of hadronic atom level shifts. The
usual picture of a hadronic atom is a particle moving in the
combined Coulomb and nuclear potential of the nucleus.(see e.g.
\cite{Deloff}). The measured level shifts are used to extract
information about the nuclear part of the potential.

An important
subfield is the study of hadronic hydrogen, which gives information
on the primary hadron nucleon interaction. According to the usual
philosophy of hadronic atoms, this interaction is imagined as a
complex potential, the properties of which are to be deduced (at
least partly) from the measured level shifts. On the other hand, the
elementary hadron nucleon interaction can be approached also from
the field-theoretical side, trying to derive it from effective
Lagrangians and then to relate it to the level shifts \cite{Meiss}.
While the field theoretical approach can be considered as more
fundamental in this case, the basic merit of the potential picture
leading to a phenomenological potential lies in its applicability in
dynamical description of more complicated $N>2$ systems in the
quantum mechanical framework. Proper field theoretical calculations
of such systems are still beyond the real possibilities.

In both cases the level shifts and the nuclear interactions are
related via the scattering length of the nuclear interaction.  The
use of scattering length seems to be natural in the
field-theoretical case since it is the zero energy scattering
amplitude and thus diagram technique can be applied for its
calculation. On the other hand, in the potential approach this
intermediate step seems to be superfluous, with present day
computational facilities the potentials can be directly and easily
related to the level shifts.

The main purpose of the paper is to show, that the approximate
deduction of the potential parameters via the widely used
Deser-formula \cite{Des}, connecting the level shift and the
scattering length can lead to substantial differences, compared to
the direct approach (this has been known before). We  argue, that
the effort of exact solution of the Schr\"{o}dinger-equation,
yielding accurate potential parameters is not greater, than that of
applying different corrections to the Deser-formula.  Also, a new
approximate relation between the level shifts and the potentials is
derived, which practically gives the exact results.

As a demonstration of the above ideas we have made calculations for
the the  kaonic hydrogen case since recently the $\bar{K}N$
interaction has attracted some renewed interest due to the possible
existence of bound $\bar{K}$-nuclear states.

\section{Formulation of the problem}

Let us  consider the $1s$ level of kaonic hydrogen. The (mo\-del)
Hamiltonian reads
\begin{equation}
H = -{\hbar^2\over{2\mu}}\Delta_{\textbf{r}} + V_s + V_c,
\end{equation}
where the strong interaction is represented by a local central
potential
\be
V_s=V_0 \, v_s(r/b)
\end{equation}
with two parameters: its strength $V_0$ and range $b$. The
attractive Coulomb potential is
\begin{equation}
V_c=-{e^2\over r}
\end{equation}
and $\mu$ is the $K^-p$ reduced mass:
\begin{equation}
\mu = 323.478\, {\rm MeV/c^2.}
\end{equation}
The Schr\"{o}dinger equation $(H-E) \, \Psi = 0$ can be transformed into the
radial equation
\be
\label{Sch}
\left(\frac{d^2}{dr^2}-q^2-\lambda \, v_s(r/b) +
\frac{2/r_0}{r}\right)\psi(r)=0,
\ee
where
$$E=-\frac{\hbar^2}{2\mu} \, q^2, \quad \lambda=\frac{2\mu}{\hbar^2} \, V_0$$
and $r_0$ is the Bohr radius
$$r_0=\frac{\hbar^2}{\mu e^2} = 83.594\, {\rm fm.} $$
The "atomic" energy unit in this case is
$$\varepsilon_0=\frac{\mu e^4}{\hbar^2}=17.226\, {\rm keV.}$$

In connection with Eq.~(\ref{Sch}) we are interested in the change of
the eigenvalue $q^2$ compared to the $1s$ eigenvalue $q_0^2$ of the pure
Coulomb equation
\be
\label{Ce}
\left(\frac{d^2}{dr^2}-q_0^2+\frac{2/r_0}{r}\right)\varphi(r) = 0.
\ee
The solution of Eq.~(\ref{Ce}) is, of course, known:
$$q_0 = {1\over r_0}, \qquad
\varphi \sim {r\over r_0} \exp\left(-{r\over r_0} \right).$$

Let us recall some important features of Eq.~(\ref{Sch}):
\begin{description}
\item[--] the strong interaction acts on the fm scale and is vanishingly small
beyond a certain distance $R$: $b \sim$ fm, $v_s(r/b)\sim\,0$ for $r>R$;
\item[--] due to absorbtion to other channels, $\lambda$ is complex, so $q$ is complex,
too, but $Re(q)>0$, so $\psi(r)\rightarrow 0$ for $r\rightarrow 0$;
\item[--] both functions $\psi(r)$ and $\varphi(r)$ are on the $r_0$ scale
($\sim 100$ fm), while the presence of $V_s$ modifies $\psi(r)$ compared
to $\varphi(r)$ only for $r<R$; thus $q^2-q_0^2$ is small, but \emph{not} due to
the smallness of $V_s$ itself, so perturbative treatment is not justified;
\item[--] outside the range of the strong potential ($r>R$) $\psi(r)$ goes over into
the asymptotically vanishing solution of the Coulomb equation
$$\psi(r)\sim F_{out}^c(q,r)=r \exp(-qr) \, U\left(1-{1\over qr_0},2;2qr\right)$$
with $U$ being the confluent hypergeometric function of the second kind~\cite{Abr}.
\item[--] within the range of $V_s$ ($r<R$) the relation $q^2\ll\, \lambda \, v_s(r/b)$
usually holds since the Coulombic eigenvalue is in the keV range,
while the nuclear potential is of MeV order; this feature allows to
approximate $\psi(r)$ in this range by the zero-energy solution
$\psi^0(r)$.
\end{description}

Multiplying Eq.~(\ref{Sch}) by $\varphi(r)$ and Eq.~(\ref{Ce}) by $\psi(r)$,
subtracting and integrating, we obtain in the usual way:
\be
\label{ex}
q^2-q_0^2 = - \frac{\int_0^R\varphi(r)\lambda \, v_s(r/b)\, \psi(r) \,dr}
{\int_0^\infty \varphi(r) \, \psi(r)\, dr}\;,
\ee
where we used the fact, that both $\psi(r)$ and $\varphi(r)$ vanish
for $r\rightarrow\infty$ and $r\rightarrow 0$.

The expression~(\ref{ex}) is exact and independent of the normalization of
$\psi(r)$ and $\varphi(r)$. The basic question is how to relate
$$\Delta E = - {\hbar^2\over 2\mu} \,(q^2-q_0^2) $$
to the properties of $V_s$.

The traditional answer to this question was given by Deser~\cite{Des}, back in 1954.
According to him
\bea
\label{Deser}
q^2-q_0^2\approx - 4 \,{a_s\over r_0^3} \qquad \textrm{or}\nonumber\\
a_D(q)=-{r_0^3\over 4} \,(q^2-q_0^2)\approx a_s\;,
\eea
where $a_s$ is the scattering length of the strong potential $V_s$ defined as
\footnote{This is the usual textbook definition of $a_s$ corresponding to
negative scattering length for weak attractive potential ($\lambda<0$). By
some reason in meson-nuclear physics the opposite sign is used.}
\be
\label{as}
a_s = \frac{\int_0^R r \lambda \, v_s(r/b) \, \psi_s^0(r) \, dr}
{1 + \int_0^R \lambda \, v_s(r/b) \, \psi_s^0(r) \,dr} =
R - \frac{\psi_s^0(R)}{{\psi_s^0}'(R)}\;,
\ee
with $\psi_s^0(r)$ being the regular [$\psi_s^0(0)=0\,,{\psi_s^0}'(0) = 1$] solution
of the zero-energy equation with $V_s$ alone:
\be
\left({d^2\over dr^2} - \lambda  \,v_s(r/b)\right)\psi_s^0(r)=0\;.
\ee
In other words, the pure strong scattering length is approximated by the Deser
scattering length $a_D(q)$, derived from the measured $\Delta E$.

It is not straightforward to relate~(\ref{Deser}) and (\ref{as}) to the exact
expression~(\ref{ex}) or to point out clearly the approximations leading from
(\ref{ex}) to (\ref{Deser}), together with criteria for their applicability
(apart from the obvious $R\ll r_0$).

Since 1954 considerable effort has been devoted to re-derivation of Deser's
result, to considering its possible improvements or corrections to it, to studying
its special cases, e.g. a strong potential with an almost or weakly bound state,
when the scattering length becomes large~\cite{Tru}--\cite{Pop}. The most relevant
improvement is the taking into account the Coulomb distortion of the zero-energy
wave function $\psi_s^0$, leading to approximation
\be
\label{CM}
a_D(q)\approx a_{sc}
\ee
instead of~(\ref{Deser}), where $a_{sc}$ is the Coulomb modified scattering length:
\be
\label{asc}
a_{sc} = \frac{\int_0^R \Phi(r) \lambda \, v_s(r/b)\, \psi_{sc}^0(r)\, dr}
{1 + \int_0^R \Theta(r) \lambda \, v_s(r/b)\, \psi_{sc}^0(r) \, dr} =
\frac{W(\Phi,\psi_{sc}^0)|_{r=R}} {W(\Theta,\psi_{sc}^0)|_{r=R}}\;.
\ee
Here $\Phi(r)$ and $\Theta(r)$ are the suitable $q\rightarrow 0$ limits of the Coulomb
scattering functions $F$ and $G$ (see~\cite{Lam}), satisfying
\be
\left({d^2\over dr^2}+ {2/r_0\over r}\right)
\left\{
\begin{tabular}{c}
$\Phi(r)$\\
$\Theta(r)$
\end{tabular}
\right\}=0
\ee
and $\psi_{sc}^0(r)$ is the regular zero-energy solution of
\be
\left({d^2\over dr^2} - \lambda \, v_s(r/b)+{2/r_0\over r}\right)\psi_{sc}^0(r)=0\;,
\ee
while $W$ denotes the wronskian of the two functions.

Now, the usual way, how~(\ref{Deser}) and~(\ref{CM}) are
used to relate a strong potential $V_s$ to the measured energy shift is to
approximate $a_s$ (or $a_{sc}$) by $a_D(q)$ and then to design a potential
$V_s$ having this  $a_s$ (or $a_{sc}$):
$$\Delta E\Rightarrow q^2-q_0^2\Rightarrow a_D(q)\Rightarrow
\begin{tabular}{c}
$a_s\approx a_D(q)$\\
$a_{sc}\approx a_D(q)$
\end{tabular}
\Rightarrow V_s \,.
$$

But due to the approximations~(\ref{Deser}) or~(\ref{CM}) the $\Delta E'$ values,
calculated with the potentials obtained in this way do not reproduce the measured
$\Delta E$-s : $\Delta E'\neq \Delta E$, so the usual experimental claim, that
measuring $\Delta E$ is equivalent to measuring $a_s$ is not valid.

However, at the present calculational level ($2007 \gg 1954$) to
find a potential, giving a prescribed $a_s$ (or $a_{sc}$) is not
easier, than to find a potential giving exactly the measured
$\Delta E$. The eigenvalue equation~(\ref{Sch}) is solved by
numerical integration in the internal region ($r\leq R$) and by
matching the logarithmic derivatives of $\psi_{in}(r)$ and the
external function $F_{out}^c(q,r)$ at $r=R$:
\be
\label{Alq}
\frac{{\psi_{in}}'(R)}{\psi_{in}(R)}-\frac{{F_{out}^c}'(q,R)}{F_{out}^c(q,R)}
= A(\lambda,q) = 0\;.
\ee
For a given range parameter $b$
the root in $q$ of Eq.~(\ref{Alq}) for fixed $\lambda$ gives the eigenvalue, while
the root in $\lambda$ for fixed $q$ yields the potential strength corresponding
to a prescribed eigenvalue. The numerical solution of Eq.~(\ref{Alq}) is
straightforward in both cases.

For demonstration of the above idea strong potentials were derived from the three
conditions
$$ \Delta E({\rm calc})=\Delta E({\rm exp}) \eqno (a) $$
$$ a_s=a_D(q_{\rm \,exp}) \eqno (b) $$
$$a_{sc}=a_D(q_{\rm \,exp})\,, \eqno (c) $$
their level shifts and strong scattering lengths were compared for four type
of commonly used potential shapes: exponential, gaussian, square well, and Yamagouchi
(non-local, separable) and different ranges $b$. The experimental $\Delta E({\rm exp})$
was taken from the KEK experiment~\cite{KEK}:
\bea\label{ad}
\nonumber
&{}&\Delta E({\rm KEK}) = -323 + 203.5i \; {\rm eV}\quad \textrm{with} \\
&{}&\quad a_D({\rm KEK}) = 0.78 - 0.49i \; {\rm fm \,.}
\eea

\section{Results and conclusions}

The main message of this work is to emphasize, that no approximations
are needed to connect the measured $\Delta E({\rm exp})$ with the properties of the
strong interaction (model) potential $V_s$. However, the desire to find a relation
between these quantities \emph{without} solving the eigenvalue equation, which is
superior to the previously used ones, motivated the derivation of another approximate
formula.

The normalization integral in the denominator of Eq. (\ref{ex}) can be
approximated as
\bea
\label{norm}
\int_0^\infty \varphi(r) \, \psi(r)\, dr \approx
\int_R^\infty \varphi(r)\, \psi(r)\, dr = \\
\nonumber
\int_R^\infty \varphi(r) \, F_{out}^c(q,r) \,dr
\eea
due to the smallness of the nuclear region $0<r<R$ compared to the whole
range of functions $\varphi(r)$ and $\psi(r)$. Using the equations satisfied by
$\varphi(r)$ and $F_{out}^c(r)$ and their vanishing for $r\rightarrow\infty$
Eq.~(\ref{norm}) can be rewritten as
\be
\label{norm1}
\int_R^\infty \varphi(r) \, F_{out}^c(q,r)\, dr =
\frac{W(F_{out}^c,\varphi)|_{r=R}}
{q^2-q_0^2} \,.
\ee
Substituting (\ref{norm1}) into (\ref{norm}) and (\ref{ex}) and, as before,
approximating $\psi(r)$ within the nuclear range by $\psi_{sc}^0(r)$ we get
finally
\be
\label{fin}
W(F_{out}^c,\varphi)|_{r=R} =
- \int_0^R \varphi(r) \lambda \, v_s(r/b) \, \psi_{sc}^0(r)\, dr\;,
\ee
which is of the form
$$w(q) = I(\lambda) \eqno (d)$$
and again can be solved either for $q$ or for $\lambda$. The results for the strong
potentials obtained from condition ($d$) are also shown in Table~1, summarizing our
results.

From these results we can make the following conclusions:
\begin{enumerate}
\item[(i)] The quality of approximations

It can be seen, that the most commonly used approximation ($b$) gives rather
poor results, in the considered case the error in $\Delta E'$ is of the order
of \mbox{$10-15 \%$}. The opposite is also true: the $a_s$-s corresponding to
potentials exactly reproducing the measured $\Delta E$ differ from the Deser
value $a_D$ approximately by the same amount. The improved approximation ($c$)
yields better strong potentials, the error both in $\Delta E'$ and $a_s$
amounts to a few \%. The best results seem to be given by approximation ($d$),
which practically reproduces the exact $\Delta E$-s and $a_s$-s of the exact
calculation ($a$).
\item[(ii)] Model independence

One of the most attractive features of the Deser formula is its model independence:
the relation between the energy shift and the scattering length is independent of
the form of the potential. This feature is confirmed by the calculations with
different potential shapes and quite different ranges. The relation between
$\Delta E$ and $a_s$ is almost potential independent, however, only within
one approximation.
\item[(iii)] Finally, to the question ,,which  feature of a given potential
instead of $a_s$ or $a_{sc}$ determines $\Delta E$~? ``, our answer
is:
$$ I(\lambda) = - \int_0^R \varphi(r) \lambda \, v_s(r/b) \,\psi_{sc}^0(r)\, dr\;$$
through the relation ($d$).
\item[(iv)] The widely used ,,measured`` values of $a_s({\rm KEK})$ for the kaonic hydrogen
should be changed from $a_D(q_{\rm exp})$ (\ref{ad}) to
$$ a_s({\rm KEK})\simeq 0.85 - 0.62i \; {\rm fm}$$
since this is the value, which corresponds to potentials exactly reproducing the
measured level shift.

The same calculation was also performed for the case of the DEAR $1s$ level
shift~\cite{DEAR},
and while the overall results are qualitatively the same as before, the correct
scattering length corresponding to the DEAR shift is
$$a_s({\rm DEAR})\simeq 0.49 - 0.35i  \; {\rm fm}$$
instead of the widely adopted $a_D({\rm DEAR}) = 0.47-0.30i$ fm.

\end{enumerate}
\begin{acknowledgments}
The work was supported by the Czech GA AVCR grant A100480617.
\end{acknowledgments}

\begin{sidewaystable}[h]
\begin{center}
\begin{minipage}[c]{6in}
\caption{Calculated kaonic hydrogen $1s$ level shifts $\Delta$E
and scattering lengths $a_s$ of different strong interaction potentials,
derived from the measured KEK level shift $\Delta E({\rm exp}) = -323 + 203.5i$ eV
using the conditions $(a)-(d)$ (see the text).}
\end{minipage}
\begin{tabular}{c cccccccc}
 \hline\noalign{\smallskip}
 {} & \multicolumn{2}{c}{Gauss} &
\multicolumn{2}{c}{Exponential} & \multicolumn{2}{c}{Square well}
& \multicolumn{2}{c}{Yamaguchi} \\
\noalign{\smallskip}\hline\noalign{\smallskip}
$b$, fm & $a_s$, fm & $\Delta$E, eV & $a_s$, fm & $\Delta$E, eV &
$a_s$, fm & $\Delta$E, eV & $a_s$, fm & $\Delta$E, eV \\
\noalign{\smallskip}\hline\noalign{\smallskip}
{} & \multicolumn{8}{c}{$V_s$ from $\Delta$E(calc)=$\Delta$E(exp)} \\
\noalign{\smallskip}\hline\noalign{\smallskip}
 1.0 & $0.85 - 0.61i$ & {} & $0.88 - 0.61i$ & {}
& $0.84 - 0.61i$ & {} & $0.86 - 0.61i$ & {} \\
 0.5 & $0.84 - 0.62i$ & $-323 + 204i$ & $0.85 - 0.61i$ & $-323 + 204i$
& $0.84 - 0.62i$ & $-323 + 204i$ & $0.84 - 0.61i$ & $-323 + 204i$ \\
 0.25 & $0.84 - 0.63i$ & {} & $0.84 - 0.62i$ & {}
& $0.84 - 0.64i$ & {} & $0.84 - 0.62i$ & {} \\
\noalign{\smallskip}\hline\noalign{\smallskip}
{} & \multicolumn{8}{c}{$V_s$ from $a_s = a_D$} \\
\noalign{\smallskip}\hline\noalign{\smallskip}
 1.0 & {} & $-298 + 168i$ & {} & $-289 + 167i$
& {} & $-299 + 167i$ & {} & $-294 + 168i$ \\
 0.5 & $0.78 - 0.49i$ & $-299 + 165i$ & $0.78 - 0.49i$ & $-296 + 168i$
& $0.78 - 0.49i$ & $-299 + 164i$ & $0.78 - 0.49i$ & $-298 + 167i$ \\
 0.25 & {} & $-299 + 162i$ & {} & $-298 + 166i$
 & {} & $-298 + 161i$ & {} & $-300 + 165i$ \\
\noalign{\smallskip}\hline\noalign{\smallskip}
{} & \multicolumn{8}{c}{$V_s$ from $a_{sc} = a_D$} \\
\noalign{\smallskip}\hline\noalign{\smallskip}
 1.0 & $0.83 - 0.57i$ & $-317 + 191i$ & $0.86 - 0.57i$ & $-317 + 191i$
& $0.83 - 0.57i$ & $-317 + 191i$ & $0.84 - 0.58i$ & $-317 + 191i$ \\
 0.5 & $0.83 - 0.58i$ & $-317 + 191i$ & $0.84 - 0.57i$ & $-317 + 191i$
& $0.83 - 0.58i$ & $-317 + 191i$ & $0.83 - 0.57i$ & $-317 + 191i$ \\
 0.25 & $0.83 - 0.59i$ & $-317 + 191i$ & $0.83 - 0.58i$ & $-317 + 191i$
& $0.83 - 0.60i$ & $-317 + 191i$ & $0.83 - 0.56i$ & $-317 + 191i$ \\
\noalign{\smallskip}\hline\noalign{\smallskip}
{} & \multicolumn{8}{c}{$V_s$ from $w(q) = I(\lambda)$} \\
\noalign{\smallskip}\hline\noalign{\smallskip}
 1.0 & $0.85 - 0.61i$ & $-323 + 204i$ & $0.86 - 0.61i$ & $-319 + 202i$
& $0.84 - 0.61i$ & $-323 + 204i$ & $0.86 - 0.61i$ & $-323 + 204i$ \\
 0.5 & $0.84 - 0.62i$ & $-323 + 204i$ & $0.86 - 0.61i$ & $-319 + 202i$
& $0.84 - 0.62i$ & $-322 + 204i$ & $0.84 - 0.61i$ & $-322 + 203i$ \\
 0.25 & $0.84 - 0.63i$ & $-323 + 204i$ & $0.85 - 0.61i$ & $-322 + 203i$
& $0.84 - 0.63i$ & $-323 + 204i$ & $0.84 - 0.62i$ & $-323 + 203i$ \\
\noalign{\smallskip}\hline
\end{tabular}
\end{center}
\end{sidewaystable}

\end{document}